\documentclass[review,authoryear,12pt]{elsarticle}

\usepackage{epsfig}

\usepackage{amssymb,amsthm,amsmath}
\newtheorem{thm}{Theorem}

\usepackage[ps2pdf,%
a4paper=true,%
breaklinks=true,%
colorlinks=true,%
pdfauthor={First Author et al.},%
pdftitle={Template for manuscripts in Advances in Space Research}%
]{hyperref}

\journal{Advances in Space Research}

\begin{document}

\begin{frontmatter}



\title{Tropospheric delay in microwave propagation in Nigeria.}


\author{Samuel T. Ogunjo \corref{cor}}
\address{Department of Physics, Federal University of Technology, Akure, Ondo State, Nigeria}
\cortext[cor]{Corresponding author}
\ead{stogunjo@futa.edu.ng}

\author{Joseph B. Dada}
\address{Department of Physics, Federal University of Technology, Akure, Ondo State, Nigeria}
\ead{babatunde.dada@elizadeuniversity.edu.ng}
\fntext[footnote1]{Elizade University, Ilara-Mokin, Ondo State, Nigeria}

\author{Sunday S. Oluyamo}
\address{Department of Physics, Federal University of Technology, Akure, Ondo State, Nigeria}
\ead{ssoluyamo@futa.edu.ng}

\author{Ibiyinka A Fuwape}
\address{Department of Physics, Federal University of Technology, Akure, Ondo State, Nigeria}
\ead{iafuwape@futa.edu.ng}
\fntext[footnote2]{Michael and Cecilia Ibru University, Ughelli, Delta State, Nigeria}

\begin{abstract}
Satellite communication systems suffer from the systematic error of tropospheric delay.  Accurate estimation of this delay is essential for communication budget and planning.  This study investigates the tropospheric delay in three Nigeria cities: Abuja, Lagos, Port-Harcourt using two different models (Saastominen and Hopfield).  Three year atmospheric data for surface pressure, relative humidity and temperature obtained at 5-mins interval were acquired from the Tropospheric Data Acquisition Network (TRODAN) archives.   Computed radio refractivity values showed distinct seasonal dependence in Abuja with low and high values during the dry and wet season respectively.  The Hopfield model predicts higher hydrostatic delay values than the Saastominen model.  In the non-hydrostatic delay, the two models converge to a single values at high temperature.  Theorems were proposed with proofs to explain the relationship observed between the two models.
\end{abstract}


\end{frontmatter}

\parindent=0.5 cm

\section{Introduction}\label{section label}
Satellite communication systems, such as the Global Positioning System, involves transmission of information using electromagnetic waves from a satellite to ground based stations through the atmosphere \citep{abdelfatah2015precise}.  The different layers of the troposphere causes delay and refraction of the signal as it passes through.  The microwave propagation through the atmosphere experience tropospheric delay due to electrically neutral of the atmosphere and it is completely independent of the signal frequency \citep{suleyman2013comparative,younes2016modeling}. The main meteorological parameters that affect propagation through the troposphere includes relative humidity, atmospheric temperature, and atmospheric pressure. The path traveled through by a propagated signal in the atmosphere contributed to the tropospheric propagation delays. Tropospheric delay is the variability effect of refractive index on radio signal traveling through the electrically-neutral atmosphere. This propagation delay is commonly determine by small changes in refractive index describe by the term called refractivity (N) \citep{adegoke2008effect}
\begin{equation}\label{eq11}
N = (n-1)\times10^6
\end{equation}
where  n is the an atmospheric varying index.

According to \citet{mendes1995zenith}, mismodeling of tropospheric delay in radio wave propagation is a significant source of error in space geodesy techniques.  This delay is computed as the difference between the path taken by the signal and the path the signal would have taken in a vacuum.  Tropospheric delay has two components - Hydrostatic (dry) delay and non-hydrostatic (wet) delay.   The dry component of tropospheric delay accounts for 80 - 90\% of the total delay.  Wet delay is used in obtaining precipitable water vapour.  The non-hydrostatic component, although a small component of the total delay, is difficult to estimate due to the high variability of atmospheric water vapour \citep{liu2017analysis}.

In tropical Africa, refractivity of signals has been studied extensively. However, there is a dearth of literature and research on tropospheric delay in the region.  This study investigates the  tropospheric delay at three cities in Nigeria using two models: Saastamoinen \citep{saastamoinen1972atmospheric} and Hopfield \citep{hopfield1969two}.  Modelling of tropospheric delay will help mitigate the influence of the atmosphere on communication systems.

\section{Methodology}
Data for this study was obtained from the Tropospheric Data Acquisition Network (TRODAN) which monitor, collect and provide real time meteorological data of the lower atmosphere which covers region from the surface of the Earth to the altitude of about 11 km from different locations across Nigeria using the Campbell Scientific Automatic Weather Station.  Daily data for temperature, relative humidity and surface pressure for three years (April 1, 2008 - March 31, 2011) recorded at 5 minutes interval were retrieved from the archives of TRODAN.  Statistics for the three locations considered in this study are presented in Table \ref{tab1}.

\begin{table}[h]
\renewcommand{\arraystretch}{1.3}
\caption{Geographical statistics of study locations. }
\vskip0.2in
\begin{center}
\small \begin{tabular}{|c|c|c|c|}\hline
Location & Latitude ($^oN$)& Longitude ($^oE$) & Altitude (m) \\ \hline
Abuja & 9.0667 & 7.4833 & 536\\ \hline
Lagos &6.4343 & 3.3226 &  7\\\hline
Port Harcourt & 4.7848 & 6.9918 & 20 \\\hline
  \end{tabular}
\end{center}
\label{tab1}
\end{table}

Temperature (T) in Kelvins and Surface pressure (P) in hPa were used to compute the Hopfield \citep{hopfield1969two} and Saastamoinen \citep{saastamoinen1972atmospheric} models.
The Hopfield Dry delay model is given by
\begin{equation}\label{hopfield_dry}
    Z_D^H = \frac{0.62291}{T^2} + 0.0023081P
\end{equation}

Hopfield wet delay model is given by
\begin{equation}\label{hopfield_wet}
    Z_W^H = 0.07402\cdot \frac{e}{T^2}\cdot H_T
\end{equation}

where $H_T$ is the height of the tropopause which is taken to be 12km.

Saastamoinen hydrostatic delay is computed
\begin{equation}\label{saas_dry}
    Z_D^S = \frac{0.0022767P}{1 - 0.00266 \cos 2\phi - 0.00028 h}
\end{equation}
where $\phi$ and $h$ are the ellipsoidal latitude and surface height above the ellipsoid in km, respectively. And the wet component given by
\begin{equation}\label{saas_wet}
    Z_W^S = 0.002277\left(\frac{1255}{T} + 0.05 \right)e
\end{equation}

$e$ is the partial water vapour pressure at the surface computed from the expression
\begin{equation}\label{watvap}
e = \frac{5854}{T^5}10^{20 - \frac{2950}{T}}
\end{equation}

Total refractivity is computed from the expression
\begin{equation}\label{refractivity}
    N = 77.6\frac{P}{T} + 6.48\frac{e}{T} + 3.75\times 10^5\frac{e}{T}
\end{equation}

\section{Results and Discussion}\label{sec:Results}
\begin{figure}[h]
\centerline{\includegraphics[width=\textwidth]{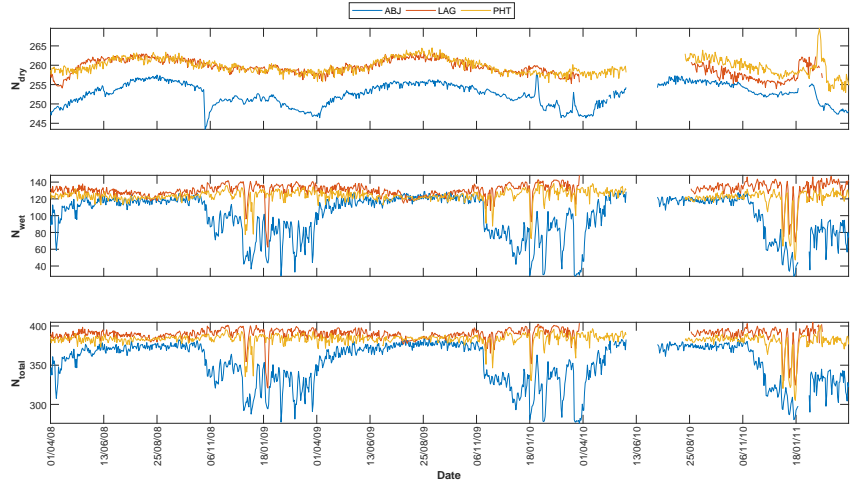}}
\caption{Temporal variation of dry component of radio refractivity (top panel), wet refractivity (middle panel) and total radio refractivity (bottom panel).}
\label{fig1}
\end{figure}

The temporal variation of dry and wet component of refractivity as well as the total refractivity over the study locations is presented in Figure \ref{fig1}.  The dry component ranges from 245 - 270 N-units and constitute the larger part of the total refractivity.   Lagos and Port Harcourt showed similar values and trend in the dry component of refractivity.  This can be attributed to the two locations being coastal regions.  Although not pronounced, Lagos and Port Harcourt also showed seasonal variations as higher values of refractivity were observed during the wet season and low values in the dry season.   Abuja, being an inland station, exhibit lower values than the coastal cities of Lagos and Port Harcourt.  This is attributed to the high value of water vapour in the atmosphere of the coastal cities.  The computed wet component of radio refractivity showed remarkable features.  During the wet season, the three stations were found to have similar values which is between 110 and 130.  However, during the dry season values of wet refractivity for Abuja dropped.    The wet component of radio refractivity was found to be a significant component of the total refractivity \citep{fuwape2016phase}.

\begin{figure}[h]
\centerline{\includegraphics[width=\textwidth]{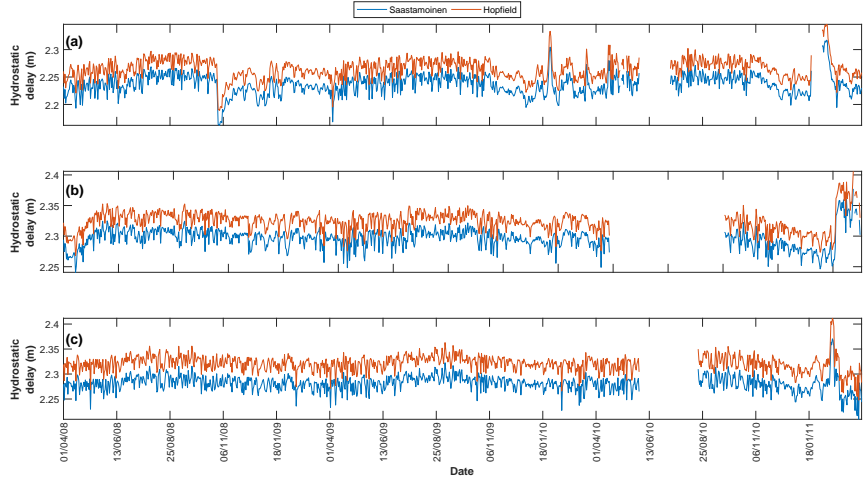}}
\caption{Temporal variation of hydrostatic zenith delay for (a) Abuja  (b) Lagos  (c) Port Harcourt.}
\label{fig2}
\end{figure}

Figure \ref{fig2} shows the temporal variation of the hydrostatic zenith delay for the three locations under consideration using the Saastomoinen and Hopfield models.  For the three locations considered, the Hopfield model suggests an higher value for hydrostatic delay than the Saastominen model, albeit, the two models were found to track each other. The difference between Hopfield and Saastomoinen hydrostatic model can be accounted for by considering Theorem \ref{thm1}. The difference between the models become smallest at high temperature.  Abuja showed the lowest hydrostatic delay compared to the coastal cities of Lagos and Port Harcourt.  This result confirms high hydrostatic delay along coastal regions and low values in inland stations as reported by \citep{fuwape2016phase}.

\begin{thm}\label{thm1}
The difference between the Saastomoinen and Hopfield hydrostatic delay can be expressed as $\frac{0.62291}{T^2} + 3.14\times 10^{-5}P$
\end{thm}

\begin{proof}
Subtracting the Hopfield hydrostatic model (Equation \ref{hopfield_dry}) from the Saastomoinen hydrostatic model (Equation \ref{saas_dry}), we obtain
\begin{equation}\label{prf1}
    Z_D^H - Z_D^S \approx \frac{0.62291}{T^2} + 3.14\times 10^{-5}P
\end{equation}
 At low altitude ($h < 1km$), $0.00028 h \approx 0$.  Also, $\max(0.00266\cos 2\phi = 0.00266)$. Hence, we assumed $1 - 0.00266\cos 2\phi - 0.00028 h \approx 1$.
\end{proof}

\begin{figure}[h]
\centerline{\includegraphics[width=\textwidth]{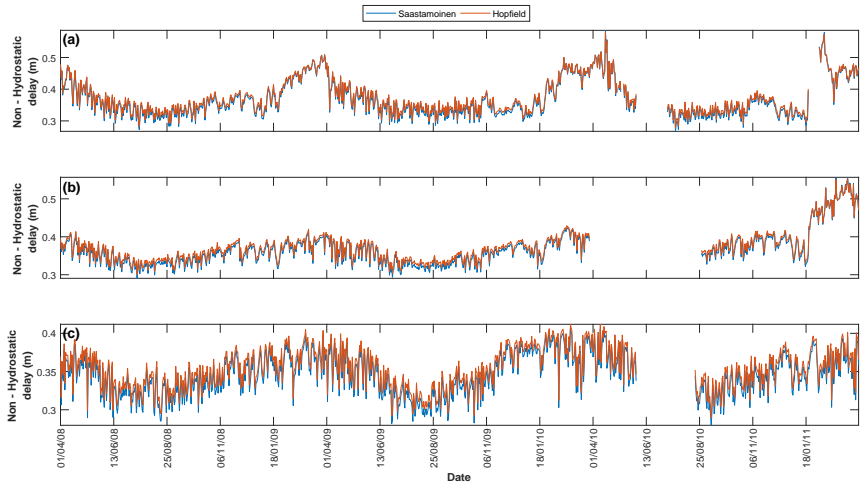}}
\caption{Temporal variation of non-hydrostatic delay for (a) Abuja  (b) Lagos  (c) Port Harcourt.}
\label{fig3}
\end{figure}

\begin{figure}[h]
\centerline{\includegraphics[width=\textwidth]{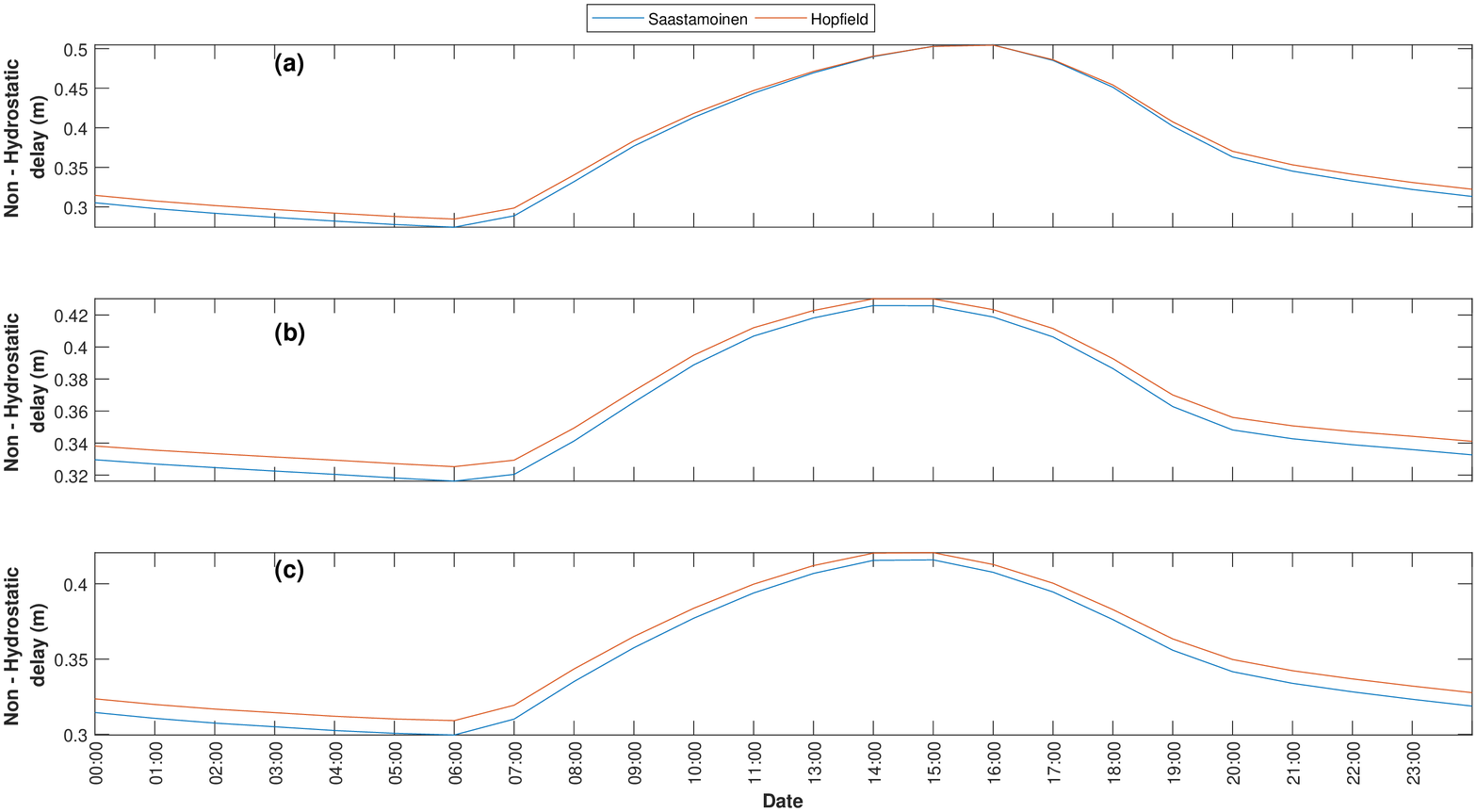}}
\caption{Diurnal variation of non-hydrostatic delay for (a) Abuja  (b) Lagos  (c) Port Harcourt.}
\label{fig4}
\end{figure}

\begin{figure}[h]
\centerline{\includegraphics[width=\textwidth]{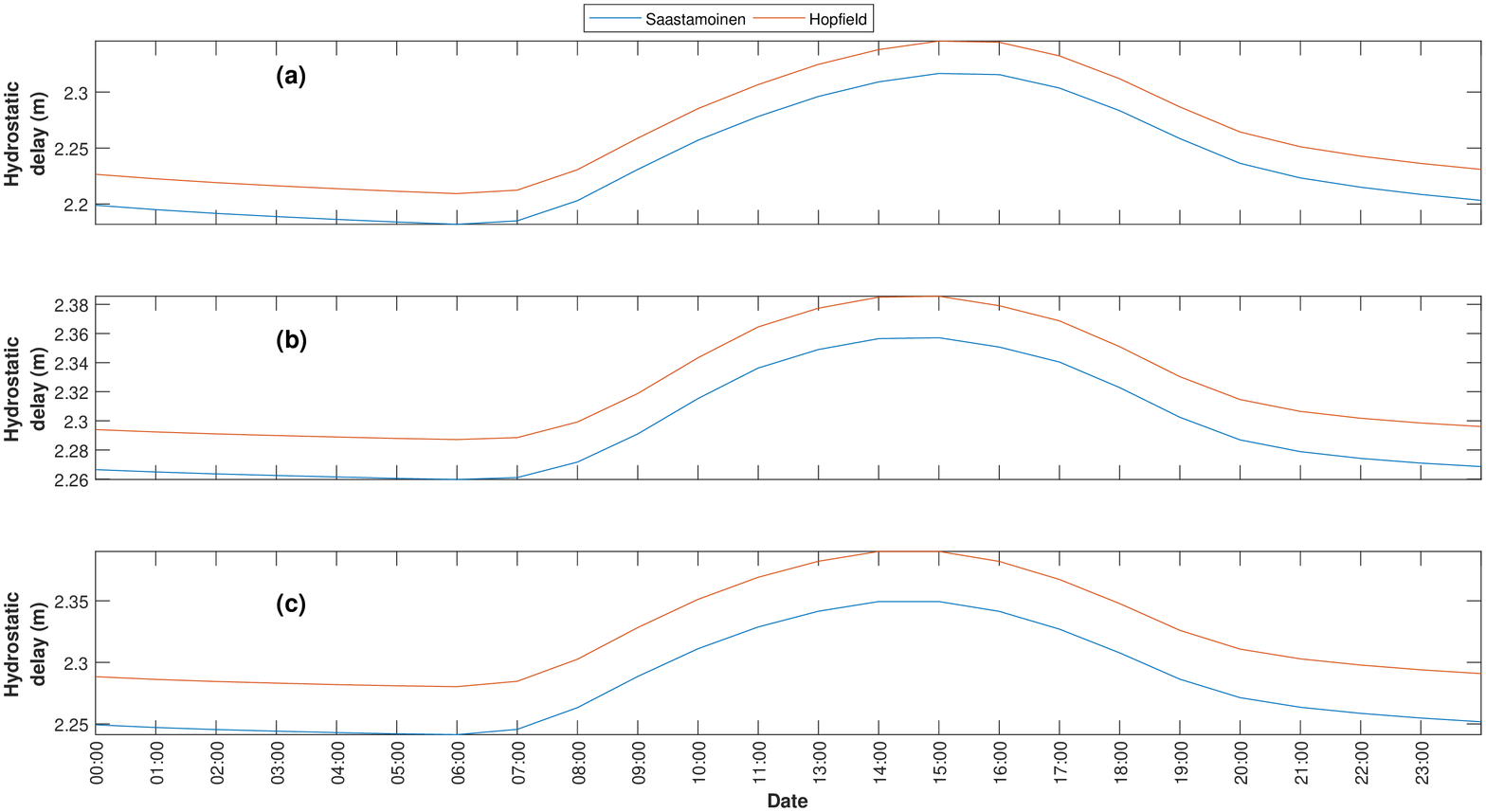}}
\caption{Diurnal variation of hydrostatic delay for (a) Abuja  (b) Lagos  (c) Port Harcourt.}
\label{fig5}
\end{figure}

Non-hydrostatic delay (Figure \ref{fig3}) showed seasonal variation.  Values of the non-hydrostatic delay were found to be low during the wet season and high during the dry season.  Values of non-hydrostatic delay were found to be closely related but not identical for Lagos and Port-Harcourt.  However, identical values were found for the two models in Abuja during the dry season in the temporal variation of non-hydrostatic delay.   The diurnal variation of non-hydrostatic and hydrostatic delay are shown in Figures \ref{fig4} and \ref{fig5} respectively.  The values of the dry and wet delays were found to rise from the low values in the morning to maximum at mid-day.  The hydrostatic and non-hydrostatic delay for Abuja was found to peak at 1500 local time while Lagos and Port-Harcourt peaked at 1400 hours local time. In Abuja, the two models converged to the same values as it approaches mid-day in the non-hydrostatic delay.  The smallest differences between the two models in the hydrostatic delay were observed in Abuja.  To account for the convergence between the Saastominen and Hopfield wet delay models during the dry season and mid-day, we propose Theorem \ref{thm2}.  During the dry season and around mid-day, the temperature in Abuja can rise beyong 307 K, hence, the convergence of the two models.

\begin{thm}\label{thm2}
Given $H_T = 12km$, the Saastominen and Hopfield non-hydrostatic delay are equivalent when temperature equals $307 K$
\end{thm}

\begin{proof}
Equating the non-hydrostatic Saastominen (Equation \ref{saas_wet}) and Hopfield (Equation \ref{hopfield_wet}), we obtain
\begin{equation}\label{prf2}
    2.857635T + 0.00011385T^2=0.07402H_T
\end{equation}
Assuming $H_T = 12km$ and solving the quadratic equation yields $T = 307K$
\end{proof}

\section{Conclusion}
In this paper, radio refractivity and tropospheric delay for three tropical locations within Nigeria have been investigated.  It was established that the coastal climate of Lagos and Port-Harcourt is responsible for the similar trend and values in the dry, wet and total refractivity.  The refractivity values for Abuja, an inland city, was found to have lower values than that of the coastal cities.  This factor is also responsible for the visible seasonal behaviour of refractivity in Abuja.   The temporal variation of hydrostatic and non-hydrostatic tropospheric delay was also studied in this research using the Saastominen and Hopfield models.  The difference between both models for the hydrostatic delay was attributed to the temperature dependence of both models.  In the non-hydrostatic delay, the two models were found to have similar values at a temperature of $33^oC$.  Proofs were presented to established the proposed theorems for hydrostatic and non-hydrostatic delays.

\section*{Acknowledgements}
The results presented in this paper rely on TRODAN data collected and managed by the Centre for Atmospheric Research, National Space Research and Development Agency, Federal Ministry of Science and Technology, Anyigba, Nigeria. We thank the Centre for Atmospheric Research and their partners for promoting high standards of atmospheric observatory practice as well as the Federal Government of Nigeria for continuous funding of the Nigerian Space programme (www.carnasrda.com).

\bibliographystyle{elsarticle-num-names}

\end{document}